\documentclass[aps,prl,superscriptaddress,reprint,floatfix]{revtex4-1}
\usepackage{amsmath}    
\usepackage{graphicx}   
\usepackage{array}
\usepackage[dvipdfx,cmyk]{xcolor}     
\usepackage[caption=false]{subfig}  
\usepackage[colorlinks,linkcolor=red,citecolor=brown,urlcolor=blue]{hyperref}
\usepackage{dsfont,bm}
\usepackage{color}

\usepackage{amsmath} 
\usepackage{amsthm} 
\usepackage{amssymb}	
\usepackage{graphicx} 

\newcommand{\abs}[1]{\left| #1 \right|} 
\newcommand{\avg}[1]{\left< #1 \right>} 

\newcommand{\ket}[1]{\left| #1 \right>} 
\newcommand{\bra}[1]{\left< #1 \right|} 
\newcommand{\proj}[1]{\left| #1 \right>\left< #1 \right|} 
\newcommand{\braket}[2]{\left< #1 \vphantom{#2} \right|
 \left. #2 \vphantom{#1} \right>} 
\let\baraccent=\= 
\renewcommand{\=}[1]{\stackrel{#1}{=}} 

\theoremstyle{definition}

\theoremstyle{remark}

\newcommand{\be}{\begin{equation}}
\newcommand{\ee}{\end{equation}}
\newcommand{\ben}{\begin{eqnarray}}
\newcommand{\een}{\end{eqnarray}}
\newcommand{\bes}{\begin{subequations}}
\newcommand{\ees}{\end{subequations}}
\newcommand{\bF}{\begin{figure}}
\newcommand{\eF}{\end{figure}}

\def \cv{\mathrm{Cov}}
\def \I#1{\mathcal{I}_{ #1}}
\def\tr#1{{\rm{Tr}}\left[#1\right]}

\begin{document}
\title{Quantum Enhanced Multiple Phase Estimation}

\author{Peter C. Humphreys}
\author{Marco Barbieri}
\author{Animesh Datta}
\author{Ian A. Walmsley}
\affiliation{Clarendon Laboratory, Department of Physics, University of Oxford, OX1 3PU, United Kingdom}

\begin{abstract}

We study the simultaneous estimation of multiple phases as a discretised model for the imaging of a phase object. We identify quantum probe states that provide an enhancement compared to the best quantum scheme for the estimation of each individual phase separately, as well as improvements over classical strategies. Our strategy provides an advantage in the variance of the estimation over individual quantum estimation schemes that scales as $\mathcal{O}(d),$ where $d$ is the number of phases. Finally, we study the attainability of this limit using realistic probes and photon-number-resolving detectors. This is a problem in which an intrinsic advantage is derived from the estimation of multiple parameters simultaneously.

\end{abstract}
\maketitle

\emph{Introduction-} Recent developments in quantum metrology point to a new frontier of parameter estimation in which exploiting quantum states enables higher precision than can be achieved using only classical resources. Much of the work in this field to date has been directed towards the estimation of a single Hamiltonian parameter. This has been explored both theoretically~\cite{Holland1993,Bollinger1996,Giovannetti2001,Giovannetti2006,Holevo1982,Braunstein1994,Lee2002,Dorner2009,Genoni2012,Datta2011,DArianoParis97,DAriano1998,Soderholm03} and experimentally, with the estimation of optical phase shifts by means of interferometry providing the dominant paradigm, in the setting of photonic systems as the leading platform~\cite{Mitchell2004,Nagata2007,Afek2010,Yonezawa2012,Vourdas2005}. 

One of the most important metrology problems to the wider research community is that of microscopy and imaging. Producing a quantum advantage in imaging would be of significant benefit in fields such as biology, particularly for the imaging of samples that are sensitive to the total illumination. Various approaches to quantum imaging have  been proposed, typically exploring methods for increasing the diffraction limited resolution of optical imaging systems~\cite{Giovannetti2009, Brainis2011a, Schwartz2012,Guerrieri2010, Brida2010, Tsang2009, Shin2011}. A recent classical investigation of quantum enhanced imaging made use of point estimation theory, quantifying differences between images by means of a single parameter~\cite{Perez-Delgado2012}. However, imaging is inherently a multi-parameter estimation problem, and deeper insights can be gained by studying it as such.

In this Letter, we consider a discretised model for phase imaging based on this approach. Phase imaging is a cornerstone of optical microscopy, typically realised using the related techniques of phase contrast and differential interference contrast imaging \cite{Preza1999}, that allows differences in refractive index to be detected in otherwise transparent media. So far, the potential for quantum enhancements to these techniques has yet to be explored. Our approach maps phase imaging onto the problem of multiple simultaneous phase estimation. 

Our results provide a strategy for the estimation of multiple phases using correlated quantum states, in which the multi-parameter nature of the problem leads to an intrinsic benefit when exploiting quantum resources. A surprising outcome of our analysis is that our quantum strategy provides an $\mathcal{O}(d)$ advantage, where $d$ is the number of phases, over the optimal quantum individual estimation scheme of using $N00N$ states~\cite{Lee2002}. We further show that a resource advantage can be provided over the best classical phase estimation schemes.\\

\emph{Phase imaging-} We adopt a discretised model of phase imaging, in which we address the question of how to estimate $d$ independent phases most efficiently with $N$ photons. We note that earlier works have explored other aspects of multiple parameter estimation from a quantum information perspective. In the case of the estimation of parameters characterising a set of non-commuting unitary operations, it was shown that entangled states and measurements can attain the Heisenberg limit in the number of photons used in each probe state~\cite{Imai2007, Fujiwara01,Ballester2004a}. In the commuting case, the problem of estimating $d$ phases with an ensemble of single-photon probe states has been considered. A Bayesian approach showed that the cost of estimation increases with the number of parameters involved~\cite{Macchiavello}, and a Fisher information based approach showed that entangling two multi-level systems provides no advantage over using a single multi-level system~\cite{Ballester2004}.  More recently, the error associated with estimating two phases using three and four mode interferometers (and three and four photons respectively) has been investigated~\cite{Spagnolo2013}.
\begin{figure}[htbp]
\begin{center}
\includegraphics[width=5cm]{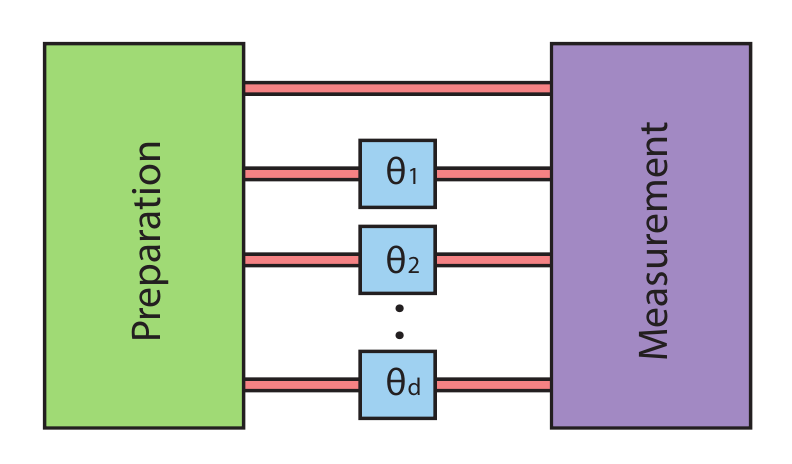}
\caption{Discretised phase imaging model. We consider the simultaneous estimation of \(d\) phases using a setup consisting of state preparation (green), independent phase application in each mode (blue) and state measurement (purple).}
\label{fig:multiPhaseConcept}
\end{center}
\end{figure}

We now turn to the general case of determining multiple independent phases by distributing N photons across a probe state in an optimal manner. Our discretised phase imaging model consists of a ${d+1}$-mode interferometer with a preparation, an interaction, and a measurement stage as in Fig.~(\ref{fig:multiPhaseConcept}). The preparation stage creates an arbitrary pure input state of the form
\be
\label{eqn:probe}
\ket{\psi} = \sum_{k=1}^D \alpha_{k} \ket{N_{k,0},N_{k,1},..N_{k,d}} \equiv \sum_{k=1}^D \alpha_{k} \ket{\mathbf{N}_k},
\ee
The distribution of photons in a given configuration $k$ is expressed compactly by a vector $\mathbf{N}_k$, composed of individual components $N_{k,m}$ that give the number of photons in mode $m$, such that $\sum_{m=0}^d N_{k,m}=N$. $D=(N+d)!/N!d!$ is the number of distinct configurations of distributing $N$ photons across $d+1$ modes. Exploiting the global phase freedom of the problem, we choose the mode labelled 0 as our reference mode and therefore the modes registering the phases are labeled $\{1,\cdots,d\}.$ To each of these configurations, we associate a complex amplitude $\alpha_k.$ The state is normalised by requiring that $\sum_{k=1}^D |\alpha_k|^2=1$.

The input state acquires a phase through the unitary transformation $U_{\bm{\theta}} = e^{i \sum_{m=1}^d \hat{N}_m \theta_m},$ where $\theta_m$ is the phase accrued and $\hat{N}_m$ the number operator for mode $m$. Denoting $\bm{\theta} = \{\theta_1,\cdots,\theta_d\},$ the evolved state is given by
\be
\label{eqn:phasestate}
\ket{\psi_{\bm{\theta}}}=U_{\bm{\theta}} \ket{\psi} = \sum_{k=1}^D \alpha_{k}  e^{\mathrm{i}\mathbf{N}_k.\bm{\theta}} \ket{\mathbf{N}_k}
\ee

The precision of the estimate of $\bm{\theta},$ governed by its covariance matrix $\cv(\bm{\theta})$, is lower bounded via the the quantum Cram\'{e}r-Rao bound (QCRB)~\cite{Helstrom1976}
\be
\label{eqn:qcrb}
\cv(\bm{\theta}) \geq(M\I{\bm{\theta}})^{-1},
\ee
where $\I{\bm{\theta}}$ is the quantum Fisher information (QFI) matrix and $M$ is the classical contribution from repeating the experiment~\footnote{In this multiparameter scenario, a single trial has to necessarily consist of an estimate of all d phases.}. This is a  $d \times d$ sized matrix inequality which is satisfied when $\cv(\bm{\theta}) - (M\I{\bm{\theta}})^{-1}$ is a positive matrix. The QFI matrix is defined as~\cite{Helstrom1976,Paris2009}
\be
[\I{\bm{\theta}}]_{\:l,m}=\frac{1}{2}\bra{\psi_{\bm{\theta}}}(L_{l}L_{m}+L_{m}L_{l})\ket{\psi_{\bm{\theta}}}.
\ee
where the operators $L_m$ are called symmetric logarithmic derivatives, defined for pure states by
\be
\label{eq:sld}
L_{\theta_l} = 2 \left(\ket{\partial_{\theta_l} \psi_{\bm{\theta}}}\bra{\psi_{\bm{\theta}}} + \ket{\psi_{\bm{\theta}}} \bra{\partial_{\theta_l} \psi_{\bm{\theta}}}\right)
\ee
We show in Supplementary Information Section~I that the QFI matrix associated with the estimation of the phases in our interferometer is
\begin{align}
\label{eqn:QFI}
 \I{\bm{\theta}} &= 4 \sum_{i} \abs{\alpha_i}^2 \mathbf{N}_i \mathbf{N}_i^T - 4 \sum_{i,j} \abs{\alpha_{i}}^2 \abs{\alpha_{j}}^2 \mathbf{N}_i \mathbf{N}_{j}^T .
\end{align}
 
For this study, we consider only the ideal case of pure states. In this case, the bound is guaranteed to be saturated if the condition $\mathrm{Im} \left [ \bra{\psi_{\bm{\theta}}}L_lL_m\ket{\psi_{\bm{\theta}}} \right ]=0$ is satisfied, which is true in our case for all $l,m$ and $\bm{\theta}$~\cite{Matsumoto2002}. Thus the QCRB can be saturated for the estimation of multiple phases simultaneously given the input states we study in Eq.~(\ref{eqn:probe}).

Since we are interested in purely quantum enhancements, we henceforth set $M=1$ in Eq.~(\ref{eqn:qcrb}). Then taking the trace of both sides gives a lower bound on the total variance of all the phases estimated
\be
\label{eq:varbound}
\abs{\Delta\bm{\theta}}^2 \equiv \sum_{m=1}^d \delta\theta^2_m \equiv \tr{\cv(\bm{\theta})} \geq \tr{\I{\bm{\theta}}^{-1}}.
\ee
The saturation of the matrix QCRB implies a saturation of the above inequality, and in the rest of this paper, we will be concerned with minimising $\abs{\Delta\bm{\theta}}^2.$\\

\emph{Optimal probe states- }It is well-known that the best quantum probe of $N$ photons for estimating a single phase is the \(\mathrm{N00N}\) state which saturates the corresponding QCRB and attains the Heisenberg limit of $\abs{\Delta\bm{\theta}}^2=1/N^2$~\cite{Lee2002}. The origin of this scaling is the number variance for the two modes, which scales as $N^2.$ Based on this intuition, we consider a generalisation in which our quantum probe is a coherent superposition of $N$ photons in one of the modes and none in any of the other $d$ modes. Due to the symmetry of our problem over the $d$ modes in which we choose to estimate the phases, we consider the quantum probe
\ben
\label{eqn:OptimState}
\ket{\psi} &= \alpha \left (\ket{0,N,..,0,0} + \ket{0,0,..,N,0} + .. +\ket{0,0,..,0,N}\right)\notag\\
& + \beta \ket{N,0,\cdots,0,0},
\een
such that $d\alpha^2+\beta^2=1.$ For these states, the QFI matrix can be found using Eq.~(\ref{eqn:QFI}). As the QFI only depends on the amplitude of $\alpha$ and $\beta$, we assume that they are real without loss of generality. Under this assumption, $\beta$ is uniquely determined by the normalisation condition and is therefore no longer an independent variable.
\be
\label{eqn:optQFI}
[\I{\bm{\theta}}]_{\:l,m} = 4N^2 \, (\delta_{l,m} \, \alpha^2 - \alpha^4 )
\ee

The minimum total variance in Eq.~(\ref{eq:varbound}) can be found by minimising $\tr{\I{\bm{\theta}}^{-1}}$ via differentiation with respect to $\alpha$,
\be
\label{eqn:ourbound}
\abs{\Delta{\boldsymbol{\theta}}_s}^2 = \frac{(1+\sqrt{d})^2\,d/4}{N^2}
\ee
for $\alpha = 1/\sqrt{d+\sqrt{d}}.$ We label this state as $\ket{\psi_s}$. This bound should now be compared to the variance of estimating the $d$ phases $\bm{\theta}$ using $d$ separate interferometers independently. Assuming for simplicity that $d$ is a factor of $N$~\footnote{For cases in which $d$ is not a factor of $N$, we show in Supplementary Information Section~III that this is a conservative estimate, since it underestimates the minimum error achievable using $N$ photons.}, the best quantum strategy uses \(\mathrm{N00N}\) states with a maximum of $N/d$ photons, with a variance of $d^2/N^2$ for each phase. Then the total variance for this approach is $\abs{\Delta{\boldsymbol{\theta}}_{\mathrm{ind}}}^2= d^3/N^2.$ In a classical strategy where the probe is restricted to uncorrelated coherent states of the form $\otimes^d_{i=1} \ket{\alpha_i}$, such that $\sum^d_{i=1} \bra{\alpha_i}\hat{N}_i\ket{\alpha_i} = N$, the total variance is $\abs{\Delta{\boldsymbol{\theta}}_{\mathrm{clas}}}^2= {d^2}/{N}.$

\begin{figure}[htbp]
\begin{center}
\includegraphics[width=6.5cm]{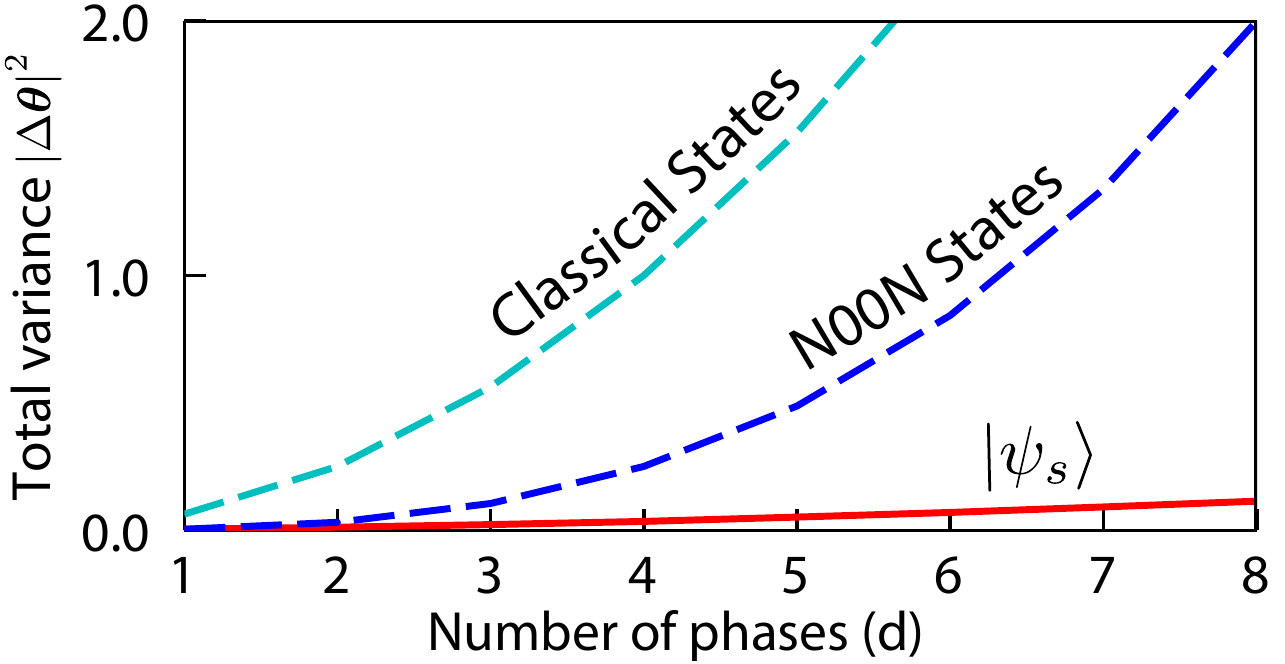}
\caption{Strategies for multiple phase estimation using $N =  16$ photons. The red line gives the total variance $\abs{\Delta{\boldsymbol{\theta}}_s}^2$ for the quantum simultaneous strategy using the states $\ket{\psi_s}$, the blue dashed line gives the variance $\abs{\Delta{\boldsymbol{\theta}}_{\mathrm{ind}}}^2$ achievable using $\mathrm{N00N}$ states, and the cyan dashed line gives the variance $\abs{\Delta{\boldsymbol{\theta}}_{\mathrm{clas}}}^2$ for an equivalent classical state.}
\label{fig:QFIScaling}
\end{center}
\end{figure}

As expected, both the quantum strategies follow the Heisenberg scaling in the total number of photons for the total variance. However, the quantum simultaneous strategy has an additional advantage over the others. Comparing the three bounds, we find
\be
\abs{\Delta{\boldsymbol{\theta}}_s}^2 \leq \abs{\Delta{\boldsymbol{\theta}}_{\mathrm{ind}}}^2 \leq \abs{\Delta{\boldsymbol{\theta}}_{\mathrm{clas}}}^2,
\ee
where the first inequality is strict for $d > 1,$ and the second for $d < N.$ As typical instances would consist of many more photons than the number of parameters to be estimated, we are guaranteed that our strategy of simultaneous quantum estimation is better than individual estimation. Furthermore, the advantage, shown in Fig~(\ref{fig:QFIScaling}), over the best quantum strategy of independent estimation improves linearly with the number of phases, scaling as $1/4d$. This is our main result.

The advantage of simultaneous quantum phase estimation is $\mathcal{O}(d)$~\footnote{The linear advantage in joint estimation provided by this quantum probe state is not unique. Setting $\alpha=\beta=1/\sqrt{d+1}$ in Eq.~(\ref{eqn:OptimState}) provides a slightly different probe state $\ket{\psi_w}$ that has a total variance of $\abs{\Delta{\boldsymbol{\theta}}}^2 = d(d+1)/2N^2,$ providing a scaling enhancement of $1/2d.$ In an actual experiment, it may be easier to produce this balanced probe state, in which the phase and reference mode amplitudes are the same. This state preserves the $\mathcal{O}(d)$ scaling improvement over individual phase estimation, and provides a total variance within a factor of 2 of that of the optimal state for large $d$}, and one might wonder if this is the maximum possible advantage that can be obtained by quantum probes of the form Eq.~(\ref{eqn:probe}). We do not have an analytic proof that this is the case, and numerical searches are hampered by the unfavourable scaling in the number of state configurations $D$, since in the limit of large $N,d,$ this is $D  \sim 2^{(N + d) S},$ where $S$ is the binary Shannon entropy of $d/(N+d).$ We have, however, performed a numerical optimisation to find the states with the minimal total variance in the parameter ranges $d=1:6, n=1:6$ and found that the optimal states always have the form in Eqn.~(\ref{eqn:OptimState}).


The definition of a trial is central for a proper accounting of resources and therefore for identifying any quantum advantages. We have defined a trial to consist of a complete characterization of all $d$ phases using $N$ photons. Alternative definitions can be considered, such as when a trial simply consists of a single illumination of the sample with $N$ photons, with freedom to use these photons differently in each trial. In the latter case, an alternative strategy of using all $N$ photons to estimate a single phase in a given trial, switching through the phases to be estimated in each trial, can also produce an $\mathcal{O}(d^2)$ scaling in the variance. Now, however,  $Nd$ photons are required in order to provide one set of estimates for the phases, this will lead to $d$ fewer trials per phase, and therefore a factor $1/d$ slower convergence to the Cram\'{e}r-Rao bound.

\emph{Optimal measurements- }We now turn to the problem of identifying measurements that can realise the quantum advantages in multi-phase estimation. Although we know that the QRCB can be saturated in principle, it is important to identify the measurements that allow us to do so in practise. In Supplementary Information Section~II, we consider positive-operator-valued measurement (POVM) sets in which one element is a projection onto the probe state after transformation by the interferometer with phases $\bm{\theta}_s$. We show that these sets saturate the QCRB at this specific point in parameter space, and that the associated classical Fisher information matrix is equal to the QFI matrix. 

One such construction, for the probe $\ket{\psi_w}$, is given by $\mathbf{\Pi}=\{\proj{\Upsilon_l}\}$ where $\ket{\Upsilon_l} = \sum_m \Upsilon_{l,m} \ket{\mathbf{N'}_m}$ and $\ket{\mathbf{N'}_m}$ is the configuration with $N$ photons in mode $m$ and no photons in any other mode. The component amplitudes are given by
\be
\Upsilon_{l,m} = \left\{
       \begin{array}{ll}
         \sqrt{\frac{(l-1)!}{(l+1)!}}, & m \leq l - 1; \\
         -\sqrt{\frac{l}{l+1}}, & m = l; \\
         0, & m > l,
       \end{array}
     \right.
\ee
for $l=1,\cdots,d$ and $m=0,\cdots,d$. The additional $l=0$ state is given by $\Upsilon_{0,m}=1/\sqrt{d+1}$. This set saturates the QFI for  $\bm{\theta}_s = \bm{0}$. An explicit construction for $d=3$ is shown in Table~I in Supplementary Information Section~II. A similar set of projectors can also be obtained for the optimal state given by Eq.~(\ref{eqn:OptimState}).

As can be seen, the probability $p_l=|\langle \psi_w\ket{\Upsilon_l}|^2$ associated with each outcome is transparently related to the phases, with $p_1$ involving only $\theta_1,$ $p_2$ only $\theta_1,\theta_2$ and so on. This suggests that an estimator could be easily created that would allow one to determine the probability distribution for the phases given a set of experimental outcomes. \\



\emph{Realistic probes and measurements- }The optimal probe states and measurements involve quantum correlated states that may be challenging to implement in practice. In this section, we present examples of probe states that may be relatively easier to prepare, and show the enhancements predicted earlier are achievable using realistic measurements.

For single parameter estimation, it was shown that the Holland-Burnett (HB) state~\cite{Holland1993,Datta2011}, generated by interfering two pure $N$ photon states on a 50/50 beam splitter, can also lead to a $1/N^2$ Heisenberg scaling in estimation. This state is significantly easier to generate than the ideal $\mathrm{N00N}$ state since it does not rely on the use of optical non-linear interactions or quantum gates. Further, these states are also known to be close to optimal with respect to losses in the quantum sensor.

We consider a multi-mode generalisation of these states, generated by means of Fourier multi-port devices that implement a quantum Fourier transform (QFT)~\cite{Vourdas2005,Lim2005,Spagnolo2013}, for two modes this is equivalent to a 50/50 beam-splitter. As in the creation of HB states, $n$ photons are input into each mode of the QFT device, leading to an $N=n(d+1)$ photon state output, that we denote $\mathrm{HB}(n,d)$. This state is then used for phase estimation. Our results include as a special case recent work by Spagnolo \emph{et al.}~\cite{Spagnolo2013} which explored the QFI associated with this device for the specific case of $d=2,3$ with $n=1$.

\begin{figure}[htbp]
\begin{center}
\includegraphics[width=7.5cm]{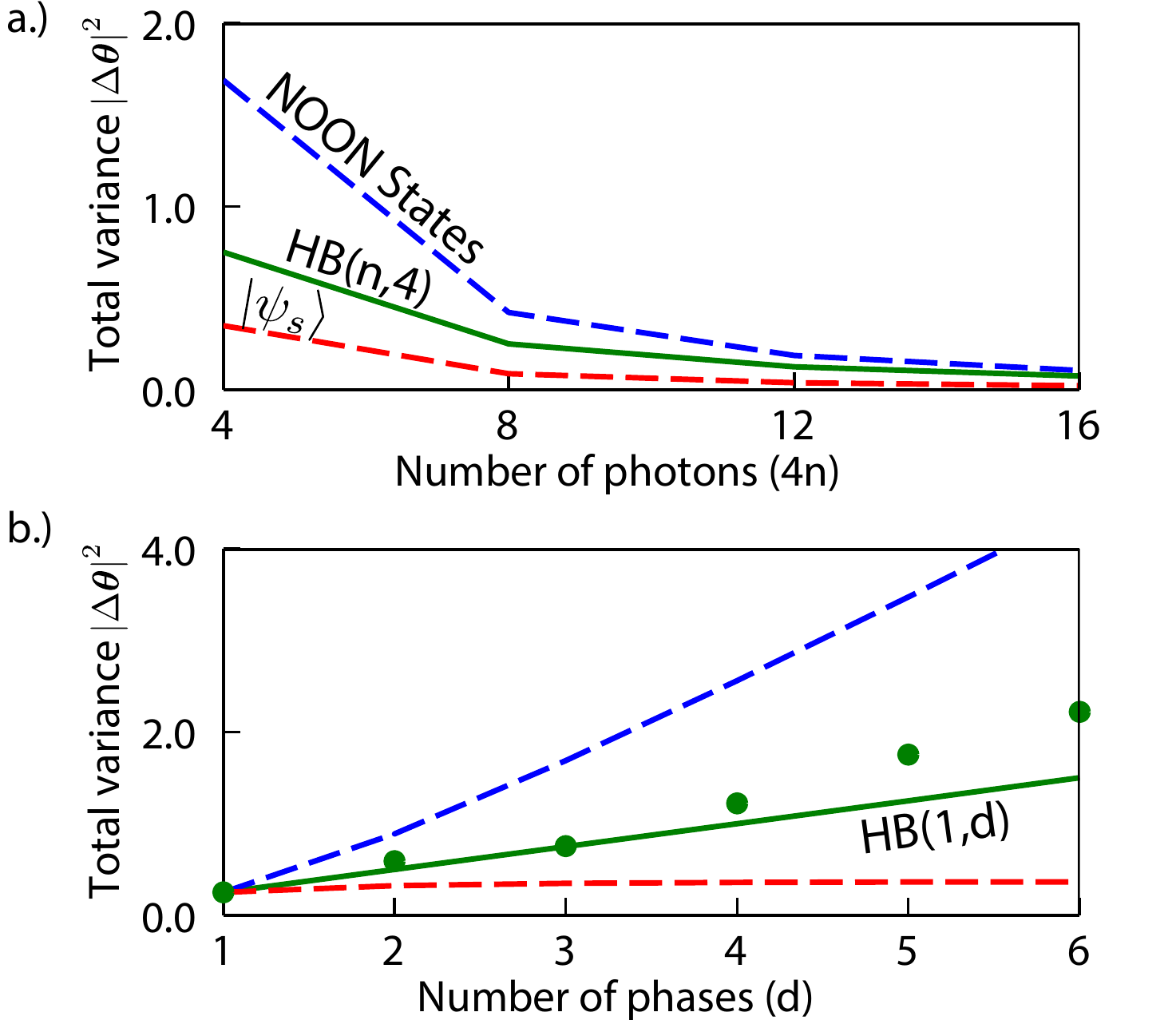}
\caption{a.) \emph{Realistic probes: }The green line gives numerical calculations of the total variance from the QCRB for the simultaneous estimation of 4 phases using $\mathrm{HB}(n,4)$ states as a function of $n$. For comparison, the blue and red dashed lines give the QCRB for equivalent $\mathrm{N00N}$ and $\ket{\psi_s}$ states respectively. b.) \emph{Realistic measurements: } The green dots show the total variance for the simultaneous estimation of $d$ phases using a $\mathrm{HB}(1,d)$ state and a measurement apparatus consisting of a Fourier multi-port followed by PNRDs. The green line gives the QCRB variance error for the same $\mathrm{HB}(1,d)$ state, while the blue and red dashed lines again give the QCRB for equivalent $\mathrm{N00N}$ and $\ket{\psi_s}$ states respectively.}
\label{fig:QFI_FI_for_HB1}
\end{center}
\end{figure}

Fig.~(\ref{fig:QFI_FI_for_HB1}) shows numerical calculations of the expected variance of estimation for these states, calculated from the QCRB (Eq.~(\ref{eq:varbound})). Our calculations suggest that the $\mathrm{HB}(1,d)$ states give the closest performance to the probe $\ket{\psi_s}$ previously considered. As the number of photons input into each mode is increased, the variance of estimation moves away from that achievable using $\ket{\psi_s},$ and approaches the error for simultaneous phase estimation using $\mathrm{N00N}$ states. The observed decrease in performance of the $\mathrm{HB}(n,d)$ state is because the probability amplitude associated with the terms in which the photons are highly bunched in one mode decreases significantly with $n$ and $d$~\cite{Tichy2012}, and it is these terms that are most sensitive to the phases in the interferometer. It is also this property, however, that ensures that these states are robust against loss in the single phase case~\cite{Datta2011}, something that is not a property of the NOON states. The degree to which multi-phase estimation can be loss-tolerant is not yet known.  

Although $\mathrm{HB}(n,d)$ states do not perform as well as comparable $\ket{\psi_s}$ probe states, they do at least as well as $\mathrm{N00N}$ states, which are just as challenging to prepare as $\ket{\psi_s}$ states. The ease of experimental generation of multi-mode HB states may make them an attractive candidate for multiple phase estimation protocols. This is particularly the case for $n=1$ states, which could be produced using heralded single photons, and demonstrate the best comparative performance over $\mathrm{N00N}$ states of the same photon number. 

In addition to the challenges of optimal state preparation, the optimal measurements involve projections onto complex multi-photon states, and thus they they may not be experimentally feasible. It is therefore important to show that an experimentally realistic measurement scheme exists that can achieve or approach the QCRB. We calculated numerically the variance of the phase estimation given by the classical Fisher information for $\mathrm{HB}(n,d)$ states using a detection scheme in which the different modes are combined using a balanced Fourier multi-port device, followed by ideal photon-number-resolving detectors (PNRD). Since the probability of different combinations of detector outcomes depends on the phases, a maximum likelihood scheme could in principle be used to estimate the phases given a set of measured detector outcomes. As the accuracy of estimation is dependent on the value of $\bm{\theta}$, numerical optimisation over the phases was used determine the minimum possible error. Calculations were carried out for the multi-mode $\mathrm{HB}(1,d)$ states (the class of $\mathrm{HB}(n,d)$  states that exhibited the best performance), and are shown in Fig.~(\ref{fig:QFI_FI_for_HB1}b). The calculated variance is comparable to the QFI, and below that achievable using $\mathrm{N00N}$ states.\\

\emph{Conclusions-} 
Our analysis of imaging as a multi-parameter estimation problem presents an alternative approach to the typical methods based on enhancing diffraction limits, and may be of interest for other quantum enhanced imaging problems.   
In addition, our results should be of wide interest as many problems, such as strain sensing, range finding and gravitational wave detection can be recast as optical phase estimation~\cite{Giovannetti2011}. They should also motivate an investigation into the nature of the quantum resources at the root of the enhancement shown.\\

\emph{Acknowledgements- }We thank J. Nunn, M.D. Vidrighin, B. Metcalf, J. Spring and W.S. Kolthammer for helpful discussions and comments on the manuscript. This work was supported by the Engineering and Physical
Sciences Research Council (EP/H03031X/1), the European Commission project
Q-ESSENCE (248095) and the Air Force
Office of Scientific Research (European Office of Aerospace
Research and Development).

\bibliography{library}
\clearpage

\onecolumngrid
\appendix

\section{Quantum Enhanced Multiple Phase Estimation Supplementary Information}

\subsection{Quantum Fisher Information}
\label{app:SLD}

Here we show how to calculate the quantum Fisher information for the estimation of multiple phases using the model described in the main text.\\

For a pure state $\ket{\psi(\bm{\theta})}$, the symmetric logarithmic derivative (SLD) for the phase $\theta_l$ is given by
\begin{align}
&L_{\theta_l}^{\text{pure}} = 2 \ket{\partial_{\theta_l} \psi(\bm{\theta})}\bra{\psi(\bm{\theta})} + 2 \ket{\psi(\bm{\theta})} \bra{\partial_{\theta_l} \psi(\bm{\theta})}
\end{align}

Using our expression for the state after the phases have been applied to it (main text Eqn.~2) we find
\begin{align}
&L_{\theta_l} = 2 \sum_{i,j} i (N_{i,l}-N_{j,l}) \: e^{i \: (\mathbf{N}_i - \mathbf{N}_j).\boldsymbol{\theta}} \alpha_{i} \: \alpha^*_{j} \ket{\mathbf{N}_i}\bra{\mathbf{N}_j}
\end{align}
In order to determine if the QCRB is achievable, we must investigate the commutativity of the SLDs. We therefore calculate
\be
L_{\theta_l}L_{\theta_m} = - 4 \sum_{i,j,k} (N_{i,l}-N_{j,l})(N_{j,m}-N_{k,m}) e^{i \: (\mathbf{N}_i - \mathbf{N}_k).\boldsymbol{\theta}} \alpha_{i} \abs{\alpha_{j}}^2 \: \alpha^*_{k} \: \ket{\mathbf{N}_i}\bra{\mathbf{N}_k}.
\ee
Since we are considering a pure state,
\be
\text{Tr}[\rho (L_{\theta_l}L_{\theta_m} )] = \bra{\psi(\boldsymbol{\theta})} L_{\theta_l}L_{\theta_m}\ket{\psi(\boldsymbol{\theta})}
= 4 \sum_{i,j,k} (N_{j,l}-N_{i,l})(N_{j,m}-N_{k,m})) \abs{\alpha_{i}}^2 \: \abs{\alpha_{j}}^2 \: \abs{\alpha_{k}}^2.
\ee
We can now multiply out the terms, rearrange these dummy indices, and sum over \(k\) to get:
\be
\text{Tr}[\rho (L_{\theta_l}L_{\theta_m} )] = 4 \sum_{i,j} (N_{i,l}N_{i,m}-N_{i,l}N_{j,m}) \abs{\alpha_{i}}^2 \: \abs{\alpha_{j}}^2
\ee
Due to the symmetry of this expression, it becomes clear that the expectation of the commutator of the SLDs is zero, which means that, in principle, it is possible to achieve the QCRB simultaneously for all phases. This can be calculated explicitly
\begin{align}
\text{Tr}[\rho (L_{\theta_l}L_{\theta_m} - L_{\theta_m}L_{\theta_l})] &= 4 \sum_{i,j} ((N_{i,l}N_{i,m}-N_{i,l}N_{j,m}) - (N_{i,m}N_{i,l}-N_{i,m}N_{j,l})) \abs{\alpha_{i}}^2 \: \abs{\alpha_{j}}^2\\
&=0\notag
\end{align}

We can also use the expression for the SLD to calculate the QFI 
\begin{align}
[\I{\bm{\theta}}]_{\:l,m}&=\frac{1}{2}\bra{\psi_{\bm{\theta}}}(L_{l}L_{m}+L_{m}L_{l})\ket{\psi_{\bm{\theta}}}\notag\\
&= 4\sum_{i,j} (N_{i,l}N_{i,m}-N_{i,l}N_{j,m}) \abs{\alpha_{i}}^2 \:  \abs{\alpha_{j}}^2\notag\\
&= 4\sum_{  i} N_{i,l} N_{i,m} \abs{\alpha_{i}}^2 - 4 \sum_{  i,j} N_{i,l}N_{j,m} \abs{\alpha_{i}}^2 \; \abs{\alpha_{j}}^2
\end{align}

This is more neatly expressed as
\begin{align}
\label{eqn:QFI}
 \I{\bm{\theta}} &= 4 \sum_{i} \abs{\alpha_i}^2 \mathbf{N}_i \mathbf{N}_i^T - 4 \sum_{i,j} \abs{\alpha_{i}}^2 \abs{\alpha_{j}}^2 \mathbf{N}_i \mathbf{N}_{j}^T .
\end{align}

It is interesting to note that, as one would expect from the single phase estimation case, for the diagonal elements \([\I{\bm{\theta}}]_{\:l,m}\) this reduces to a measurement of the variance in the occupation of mode \(l\):

\begin{align}
[\I{\bm{\theta}}]_{\:l,l} &= 4 \sum_{i} \; N_{i,l}^{\:2} \abs{\alpha_{i}}^2 - 4 \: (\sum_{i} N_{i,l} \abs{\alpha_{i}}^2)^{\:2}\\
&= 4 (\avg{N_{l}^2} -  \avg{N_{l}}^2)
\end{align}

\section{Optimal measurement to saturate the QCRB}
\label{app:OptimMeas}

It is possible to evaluate the performance of different measurement strategies through the calculation of the Fisher Information (FI) matrix~\cite{Braunstein1994}
\begin{equation}
[F_{\bm{\theta}}]_{\:l,m} = \sum_{k} \frac{\partial_{\theta_l}{p(k | \boldsymbol{\theta})\,\partial_{\theta_m}{p(k | \boldsymbol{\theta})}}}{p(k | \boldsymbol{\theta})},
\label{eqn:FImatrix}
\end{equation}
a quantitative measure of the information available for a given probe state and \emph{a specific set of measurements}. In this equation, \(p(k | \boldsymbol{\theta})\) represents the probability of obtaining the outcome associated with the positive-operator valued measure (POVM) element \(k\), given the set of interferometer phases $\bm{\theta}$. It is known that the single parameter QCRB, and the multi-parameter QCRB in a pure-state model can always be saturated\cite{Braunstein1994,Matsumoto2002}. In the former case, the optimal measurements are given by the SLDs, while in the latter case, Matsumoto~\cite{Matsumoto2002} presented a POVM with $d+2$ projectors that attains the QCRB. Here we present another method for finding a set of POVM elements that saturates the QCRB at a specific point in the space of \(\boldsymbol{\theta}\). This requires a POVM set in which one element is a projector onto the state \(\ket{\psi_s}\) associated with \emph{the arbitrarily chosen, but specific, \(\boldsymbol{\theta}_s\)} at which we want the QCRB to be saturated.\\

The only requirement on the set of POVM elements is that it must be complete. We chose to construct POVM elements using an iterative construction process in which, at each stage, we used the minimal number of basis states necessary to ensure the resulting element was orthogonal to all of the previous elements. This produced a set in which the relationship between the phases and the outcome probabilities was comparatively transparent (as shown in Table~\ref{tab:ProjTable}). To distinguish the initial POVM element associated with the probe eigenstate from the others, we label it as the \(k=1\) element. \(\{\hat{\Pi}_k \} \) refers to the complete POVM set, but the other elements will be later be denoted as \(\ket{\beta_k}\bra{\beta_k}\), so that \(\sum_{k\neq 1} \ket{\beta_k}\bra{\beta_k} + \ket{\psi_s}\bra{\psi_s} \equiv \mathds{1} \). \\

We first note that, for a pure state, the quantum Fisher information is
\begin{align}
[\I{\bm{\theta}}]_{\:l,m}  = 4 \, \mathrm{Re} \,[\braket{\partial_{\theta_l} \psi}{\partial_{\theta_m} \psi} - \braket{\partial_{\theta_l}}{\psi}\braket{\psi}{\partial_{\theta_m}}]\label{eqn:appQFI}
\end{align}

For a measurement using a given set of POVM elements,  \(\{\hat{\Pi_k} \} \), the corresponding classical Fisher information is given by
\begin{align}
&[F_{\bm{\theta}}]_{\:l,m}  = \sum_{k} \frac{\partial_{\theta_l}p(k | \boldsymbol{\theta})\,\partial_{\theta_m}p(k | \boldsymbol{\theta})}{p(k | \boldsymbol{\theta})}\\
&= \sum_{k} \frac{\partial_{\theta_l}\bra{\psi}\hat{\Pi}_k\ket{\psi}\,\partial_{\theta_m}\bra{\psi}\hat{\Pi}_k\ket{\psi}}{\bra{\psi}\hat{\Pi}_k\ket{\psi}}\\
&= \sum_{k} \frac{(\bra{\partial_{\theta_l}\psi}\hat{\Pi}_k\ket{\psi} + \bra{\psi}\hat{\Pi}_k\ket{\partial_{\theta_l}\psi})(\bra{\partial_{\theta_m}\psi}\hat{\Pi}_k\ket{\psi} + \bra{\psi}\hat{\Pi}_k\ket{\partial_{\theta_m}\psi})}{\bra{\psi}\hat{\Pi}_k\ket{\psi}}
\end{align}

Since \(\{\hat{\Pi}_k \} \) are Hermitian, this can be simplified to
\begin{align}
[F_{\bm{\theta}}]_{\:l,m}  &= \sum_{k} \frac{4\mathrm{Re}[\bra{\partial_{\theta_l}\psi}\hat{\Pi}_k\ket{\psi}]\,\mathrm{Re}[\bra{\psi}\hat{\Pi}_k\ket{\partial_{\theta_m}\psi}]}{\bra{\psi}\hat{\Pi}_k\ket{\psi}}
\end{align}
We calculate the FI available when our probe state is transformed by the set of phases \(\boldsymbol{\theta}_s\), so that it becomes \( \ket{\psi_s} \).  It should be noted that, for the remainder of the proof, $\ket{\partial_{\theta_l}\psi}$ and similar expressions will refer to their value evaluated at $\boldsymbol{\theta}_s$, and will therefore be denoted $\ket{\partial_{\theta_l}\psi_s}$.\\

We first consider the component of the FI contributed by the element  \(\ket{\psi_s}\bra{\psi_s}\). Its contribution is simply given by
\begin{align}
& \frac{4\,\mathrm{Re}[\braket{\partial_{\theta_l}\psi_s}{\psi_s}]\,\mathrm{Re}[\braket{\psi_s}{\partial_{\theta_m}\psi_s}]}{1}
\end{align}
It is easy to show that \(\mathrm{Re}[\braket{\partial_{\theta_m}\psi_s}{\psi_s}] = 0\) for any parameter $\theta_l$ \cite{Braunstein1995}, meaning that (for these specific phases), the above expression is zero. In order to evaluate the contributions of the other POVM elements as the phase tends to \(\boldsymbol{\theta}_s\), we are forced to consider their limiting values, since the denominator in the expression for \(\mathrm{F}_{lm}\) tends to zero for the contribution of these elements. We therefore evaluate \([F_{\bm{\theta}}]_{\:l,m}\)  for our probe state when the system phases are displaced from \(\boldsymbol{\theta}_s\) by a small change in the phase \(\delta\theta_j\) in an arbitrarily chosen mode \(j\). This allow us to express our state to first order as \( \ket{\psi} = \ket{\psi_s} + \delta\theta_j \ket{\partial_{\theta_j}\psi_s} \). We also note that, for these POVM elements, \(\hat{\Pi}_k\ket{\psi_s} = 0\). Expanding our expression for the FI, we find

\begin{align}
[F_{\bm{\theta_s}}]_{\:l,m} &= \sum_{k\neq 1} \frac{4 \, \delta\theta_j^2 \, \mathrm{Re}[\bra{\partial_{\theta_l}\psi_s}\hat{\Pi}_k \ket{\partial_{\theta_j}\psi_s}]\,\mathrm{Re}[\bra{\partial_{\theta_j}\psi_s} \hat{\Pi}_k\ket{\partial_{\theta_m}\psi_s}]}{\delta\theta_j^2 \bra{\partial_{\theta_j}\psi_s}\hat{\Pi}_k\ket{\partial_{\theta_j}\psi_s}}
\end{align}

The limiting expression for the elements of the FI at the point \(\boldsymbol{\theta}_s\) should be independent of the direction in which we have expanded our state to calculate this. We are therefore free to choose a convenient direction for each element independently. We choose \(j = l\) or \(j = m\), which significantly simplifies our expression for the FI.

\begin{align}
[F_{\bm{\theta_s}}]_{\:l,m} &= \sum_{k\neq 1} \frac{4 \, \mathrm{Re}[\bra{\partial_{\theta_l}\psi_s}\hat{\Pi}_k \ket{\partial_{\theta_m}\psi_s}]\,\mathrm{Re}[\bra{\partial_{\theta_m}\psi_s} \hat{\Pi}_k\ket{\partial_{\theta_m}\psi_s}]}{\bra{\partial_{\theta_m}\psi_s}\hat{\Pi}_k\ket{\partial_{\theta_m}\psi_s}}
\end{align}


Since \(\bra{\partial_{\theta_m}\psi_s}\hat{\Pi}_k\ket{\partial_{\theta_m}\psi_s}\) is by definition real, we find

\begin{align}
[F_{\bm{\theta_s}}]_{\:l,m}&= \sum_{k\neq 1} 4 \, \mathrm{Re}[\bra{\partial_{\theta_l}\psi_s}\hat{\Pi}_k \ket{\partial_{\theta_m}\psi_s}]
\end{align}

%

Using \(\sum_{k \neq 1} \ket{\beta_k}\bra{\beta_k} = \mathds{1} - \ket{\psi_s}\bra{\psi_s}\), this gives

\begin{align}
[F_{\bm{\theta_s}}]_{\:l,m} &= 4 \,\mathrm{Re} [\braket{\partial_{\theta_l}\psi_s}{\partial_{\theta_m}\psi_s} - \braket{\partial_{\theta_l}\psi_s}{\psi_s}\braket{\psi_s}{\partial_{\theta_m}\psi_s}]
\end{align}

Comparing this with Eqn.~\ref{eqn:appQFI}, we find that the expressions for the quantum and classical Fisher information matrices, for the specific set of phases \(\boldsymbol{\theta}_s\), are the same. Since this is the case, as long as the condition on the commutivity of the SLDs is also satisfied, these elements must be capable of saturating the QCRB at this point.\\

\begin{table}[h!]
\caption{Optimal measurement projectors $\ket{\Upsilon_l}$ onto the \(d+1\) basis state components of the balanced $\ket{\psi_w}$ and optimal $\ket{\psi_s}$ states respectively, with $\Upsilon_{l,m}$ as defined in the main text. The first mode is the reference mode. Shown here are the projectors for measuring \(d=3\) phases, constructed so that the measurements saturate the QCRB when $\bm{\theta} = [0,0,\dots,0]$.}
\subfloat[Balanced W state]{
\setlength{\extrarowheight}{2.5mm}
\(
\begin{array}{|c|c|c|c|cc|}
\hline
l & \Upsilon_{l,1} & \Upsilon_{l,2} & \Upsilon_{l,3} & \Upsilon_{l,4} &\\[0.4ex]\hline
1& -\frac{1}{\sqrt{2}} & \frac{1}{\sqrt{2}} & 0 & 0 &\\[1.2ex]\hline
2& -\frac{1}{\sqrt{6}} & -\frac{1}{\sqrt{6}} & \sqrt{\frac{2}{3}} & 0  &\\[1.2ex]\hline
3& -\frac{1}{2 \sqrt{3}} & -\frac{1}{2 \sqrt{3}} & -\frac{1}{2 \sqrt{3}} & \frac{\sqrt{3}}{2}  &\\[1.2ex]\hline
4& \frac{1}{2} & \frac{1}{2} & \frac{1}{2} & \frac{1}{2}  &\\[1.2ex]\hline
 \end{array}
\)
}
\subfloat[Optimal State]{
\setlength{\extrarowheight}{2.5mm}
$
\begin{array}{|c|c|c|cc|}
\hline
 \Upsilon_{l,1} & \Upsilon_{l,2} & \Upsilon_{l,3} & \Upsilon_{l,4} &\\[0.4ex]\hline
 -\sqrt{\frac{\sqrt{3}-1}{2}} & \sqrt{\frac{3-\sqrt{3}}{2}} & 0 & 0 & \\[1.2ex]\hline
-\frac{3^{3/4} \sqrt{\sqrt{3}-1}}{3+\sqrt{3}} & -\frac{\sqrt{3 \left(\sqrt{3}-1\right)}}{3+\sqrt{3}} & \sqrt{\sqrt{3}-1} & 0 & \\[1.2ex]\hline
  \sqrt{\frac{1}{2} \left(3\sqrt{3}-5\right)} & \sqrt{\frac{1}{6} \left(9-5 \sqrt{3}\right)}& \sqrt{\frac{1}{6} \left(9-5 \sqrt{3}\right)} & -\sqrt{\frac{1}{6} \left(3+\sqrt{3}\right)}  & \\[1.2ex]\hline
 \frac{1}{\sqrt{1+\sqrt{3}}} & \frac{1}{\sqrt{3+\sqrt{3}}} & \frac{1}{\sqrt{3+\sqrt{3}}} & \frac{1}{\sqrt{3+\sqrt{3}}}  & \\[1.2ex]\hline
\end{array}
$
}
\label{tab:ProjTable}
\end{table}

The saturation of the QFI stems from the strong contribution to the Fisher information of outcomes with low probabilities, since the \emph{relative change} in the frequency of these outcomes can be much larger than for outcomes with high probability. As $\bm{\theta}$ tends to $\bm{\theta}_s$, the $\ket{\psi_{s}}\bra{\psi_{s}}$ element has an outcome probability that tends to unity, while the other POVM elements give vanishing probabilities associated with their measurement outcomes. This makes the outcome distribution especially sensitive to small displacements from $\bm{\theta}_s$. There is however a caveat to this: Although the proof only requires the POVM set to consist of a minimum of two elements, one expects that $d$ linearly independent measurement outcomes are needed to estimate $d$ parameters. For sets with fewer elements than this, there is not enough information available to uniquely determine the values of each phase. Instead, a hypersurface is defined within the parameter space upon which each point is consistent with the measured set of outcomes. The position of this hypersurface can be known very accurately (due to the high precision implied by the QFI), but it will be impossible to determine where on the hypersurface the system is. With sufficient POVM elements, this degeneracy can be lifted so that this hypersurface is reduced to a unique point in the parameter space.


\section{Optimal strategy for individual phase estimation using N00N states}
\label{app:NOON}

We would like to determine the strategy that achieves a minimal total error when estimating multiple phases using N00N states to estimate each phase individually.\\

We first determine the best strategy for the estimation of two phases with an even total number of photons  \(2n\), using \(n-x\) and \(n+x\) photons for each mode respectively.
Since, for a N00N state, the variance for estimating a single phase using \(m\) photons is \(1/m^2\), we find that the total variance for estimating the two independent phases is
\begin{align}
\abs{\Delta\bm{\theta}} = \frac{1}{(n+x)^2} + \frac{1}{(n-x)^2}
\end{align}
This has a minimum when
\begin{align}
\partial_x \abs{\Delta\bm{\theta}} = -2\left(\frac{1}{(n+x)^3} - \frac{1}{(n-x)^3}\right) &= 0  \nonumber\\
\Rightarrow 3xn^2 + x^3 &= 0
\end{align}
The only real root occurs when \(x=0\). Therefore the minimum error is achieved when the number of photons used to estimate each phase is the same (\(n\)).\\

For an odd number \(2n+1\) of photons, using \(n-x\) and \(n+x+1\) photons for each mode respectively, we find
\begin{align}
\abs{\Delta\bm{\theta}} = \frac{1}{(n+x+1)^2} + \frac{1}{(n-x)^2}
\end{align}
This has a minimum when
\begin{align}
\partial_x \abs{\Delta\bm{\theta}} = -2\left(\frac{1}{(n+x+1)^3} - \frac{1}{(n-x)^3}\right) &= 0 \nonumber\\
\Rightarrow (1+2x)(x^2+x+3n^2+3n+1) &= 0
\end{align}
This only has a real root for \(x=-1/2\). Therefore the minimum realisable error is achieved when using \(n\) and \(n+1\) photons or, symmetrically, \(n+1\) and \(n\) photons to estimate the respective phases.\\

Since this is the case for the estimation of two phases, it can be seen that a similar pairwise comparison could be carried out for different phases in a \(d>2\) phase estimation problem. In each case, if there is an imbalance of more than 1 photon between the resources employed to estimate each phase, a smaller error can be achieved by rebalancing the resources between the phases. This will eventually lead to the optimal phase estimation strategy in which the photons are divided as evenly as possible between all of the phases. More specifically, if \(N\) photons are used, we define \(n\) as the quotient of \(N/d\) and \(r\) as the remainder. The best strategy employs \(n+1\) photons to estimate \(r\) phases, and \(n\) photons to estimate the other \(d-r\) phases. Therefore
$$
\abs{\Delta\bm{\theta}} = \frac{d-r}{n^2} + \frac{r}{(n+1)^2}.
$$
Here we show that the approximate expression given in the main paper for the error achievable using N00N states (denoted here $\abs{\Delta\bm{\theta}_\mathrm{approx}}$) always gives a better than (or equal to) estimate of the error achievable using N00N states as a full derivation.
$$
\abs{\Delta\bm{\theta}_\mathrm{approx}} = d \left(\frac{d}{N}\right)^2 = d \left(\frac{d}{nd+r}\right)^2.
$$
Consider
\begin{align}
\abs{\Delta\bm{\theta}_\mathrm{approx}} - \abs{\Delta\bm{\theta}} & \propto d^3 n^2 (n+1)^2 - (nd+r)^2((n+1)^2(d-r)+rn^2)\\\nonumber
&= (r-d)(r+2nr+2nd+3n^2d)\\\nonumber
& < 0,
\end{align}
since $r < d.$ Thus the approximate expression gives a better error than is actually achievable, with equality if and only if $N$ is exactly divisible by $d$.

\end{document}